# Novel Long-Wave Near-Infrared Fluorescence Bands in Tricarbocyanine Dyes

*Nitzan Dar and Rinat Ankri*[*]

Department of Physics, Faculty of Natural Science, Ariel University, Ariel 40700, Israel

Corresponding author: rinatsel@ariel.ac.il

**Abstract**

Long-wave near-infrared (LWNIR) dyes have garnered significant attention, particularly in biomedical applications, due to their ability to enhance light absorption, making them highly effective for in vivo imaging and phototherapy. Among these dyes, cyanines are notable for their broad tunability across the ultraviolet (UV) to LWNIR spectrum and their ability to form J-aggregates, which result in narrow absorption and enhanced emission peaks, often accompanied by a red-shift in their spectra. In this study, we investigate the fluorescence properties of three known tricarbocyanine dyes, uncovering new emission bands in the LWNIR region. These dyes exhibit two distinct fluorescence peaks between 1605 and 1840 nm, with an emission tail extending up to ~2200 nm. The intensity of these peaks varies depending on dye concentration. Furthermore, we assess the photostability, pH sensitivity, and thermal stability of the dyes, providing key insights into their potential for stable and efficient biomedical applications. Our study provides a deep investigation of the spectral characteristics of these dyes, seeking to enhance their potential application in biomedical imaging and phototherapy.

**Introduction**

Cyanine dyes are a fascinating topic for both fundamental and applied science. Their structure is based on a polymethine chain of various lengths. These chains contain a single positively charged nitrogen atom, usually as part of a heterocycle, which is surrounded by a counter anion[1-3]. Consequently, these molecules have a high absorption coefficient and high fluorescence emission intensity. Their absorbance and emission can be shifted in a broad spectral range from the ultraviolet to the long wavelength near infrared (LWNIR)[2,3] depending on the polymethine chain length and its terminal substituents[5]. They own tendency to self-assemble into different topologies including dimers, single and double-walled nanotubes, bundles, and sheets in a form of J-aggregates[2], which results in much narrower absorption and emission peaks, along with a redshift in their spectra compared to cyanine dye monomers. Due to their unique spectroscopic properties, cyanine dyes have multiple applications, such as in solar cells as photovoltaic layers[6] and light harvesting[7]; other applications are chemical sensors[8] shortwave infrared emitters[9], and in vivo imaging[10, 11].

Tricarbocyanines (heptamethine) are a class of cyanine dyes with a three-carbon backbone. Tricarbocyanines with a fluorescence in the range of 1500–2200 nm in the LWNIR can be used for many applications such as biomedical imaging or telecommunication. As a result, organic LWNIR emitters have attracted widespread attention due to their own advantages, including relatively smaller molecular weight, easier functionalization, shorter retention time in organisms, and more[12, 13]. For example, the indocyanine green, which emits at 830 nm, is in wide use in clinical applications, and the IR dye 800CW, which emits at 810 nm, is utilized in preclinical studies[14]. Semiconductive polymers nanoparticles in the LWNIR were also used for bioimaging and cancer detection with emission in a range of up to 1400 nm[15-17].

Multiple studies have been carried out on molecular dyes that emit in the 1100 to 1550 nm range and reach the edge of the LWNIR region[18-21]. However, despite the extensive work in this field, it is difficult to extend both the absorption and emission wavelengths of organic emitters beyond 1550 nm by structural modification. Only a few examples of tail emissions beyond 1550 nm have been reported. As far as we are aware, molecular emitters in this region with emission peaks in the LWNIR have not yet been reported. The uniqueness of this emission could be exploited for various applications such as biosensing or optical communication.

Biomedical applications can benefit from the high degree of photon scattering, which is inversely proportional to the wavelength of light, as the ratio varies depending on the tissue, as well as from the greater imaging depth (~3 cm), good temporal (20 ms) and spatial (2 μm) image resolution, better real-time monitoring capability and higher signal-to-noise ratio[22]. Also, minimal photon scattering and almost zero tissue auto-fluorescence are essential for biomedical applications. Due to their low attenuation coefficients in the LWNIR, especially in the telecommunication bands between 1530–1625 nm, they can also be used for fiber optic communication. They offer the possibility of creating long-range sensing networks that can be used in a variety of biological and chemical sensors[23].

In this article, we report the discovery of new fluorescence emission bands for the dyes IR 820, IR 806, and IR 783, referred to as VGE, NCD, and CGC, respectively (the names were chosen arbitrary), in both the NIR and LWNIR regions. We investigate the spectral properties of aqueous and ethanolic solutions of these tricarbocyanines, revealing intense and sharp emission bands with a large Stokes shift (805-1085~ nm; 6450-7300~ $cm^{-1}$) relative to their absorption peaks. For the first time, we demonstrate that these dyes exhibit two distinct emission bands within the 1590-1880 nm range, along with a fluorescence emission tail extending to ~2200 nm. Additionally, we explore the structural arrangement of the tricarbocyanines, nuclear magnetic resonance (NMR), and dynamic light scattering (DLS) of these solutions. Stability tests were conducted under various conditions, including irradiation, immersion in water at different pH levels, and heat exposure. Fluorescence lifetimes for CGC and NCD were also measured, providing further insight into their photophysical behavior.

**Experimental details**

**VGE, NCD and CGC Preparation**

IR 820 was purchased from Angene (London, England), and dissolved in ethanol AR (Bio-Lab, Israel) to form a 0.253 µM stock solution. IR 783 and IR 806 were purchased from (Merck, Israel). and dissolved in ethanol to form 1.27 µM and 0.74 µM stock solutions, respectively. For each experiment, the ethanolic solution was diluted in one of the following solvents: double distilled (dd) water, ethanol AR (Bio-Lab, Israel) solutions, buffer 4,7 and 10 aqueous solutions (Certipur, Canada).

**Absorption, Emission and Fluorescence Lifetime Measurements**

Absorbance spectra were measured with a FP-8500 spectrofluorometer (Jasco, Japan), and the fluorescence emission, fluorescence lifetime and photostability spectra were measured using the Fluorolog-Quanta Master (Horiba scientific, Japan) spectrophotometer with a 100 watt tungsten lamp. Data was analyzed using the FelixFL software (version 1.0.33.0, Horiba scientific, Japan). Mostly, 1 cm optical path lengths quartz cuvettes were used for both absorbance and fluorescence measurements. 7-30 µM were the typical concentrations for fluorescence measurements, unless otherwise stated, integration times were 1 second and a step size of 1 nm at the wavelength. The experiments were carried out at ambient temperature and the emission peaks were corrected with excitation correction spectra. Both the width of the excitation and emission slit were adjusted for optimal emission (1-6 million units). Fluorescence lifetime measurements (FLT) were performed using the time correlated single photon counting (TCSPC) method, with a thermally cooled Hamamatsu LWNIR photon multiplier tube photon detector. The lifetime measurements were carried out using a delta diode of $830 \pm 10$ nm, with an extremely narrow 50 ps pulse width, a 0.6 mW average power, and a 100 MHz repetition rate. The fluorescence decay curves were analyzed using the FelixFL decay analysis software, based on a multiexponential model which involves an iterative reconvolution process.

**Dynamic Light Scattering (DLS) and NMR measurements**

Size distribution measurements of the aggregates were carried out using the dynamic light scattering (DLS) system Litesizer™ 500 Particle Analyzer (Anton Paar, AUT) at ambient temperature. Concentrations of analyte were 1.27, 0.74, 0.1 or 0.01 µM. Calculations were done using the refractive index of the solvent with 10 sec for each cycle and 60 cycles per measurement. $^1$H NMR of 0.013 and 1.27 mM of CGC in of D$_2$O (Andover, MA) were recorded using a 400 MHz NMR spectrometer (Bruker BioSpin, Rheinstetten, Germany), chemical shifts were reported in parts per million (ppm) units.

**Results and Discussion**

**Absorption and Fluorescence of VGE**

The spectral bands of the tricarbocyanines VGE, NCD and CGC were first investigated. Based on our previous paper[24], we chose dye concentrations of 20 µM, which show high signal-to-noise ratio (SNR) absorption and emission spectra.

Absorption spectra of 25 µM VGE in ethanol showed a peak at 821 nm, with a shoulder appearing around 730 nm (Figure 1a), both peaks were redshifted compared to the absorption in water[24]. Building upon the findings from our previous work[24], we expanded our investigation of VGE by measuring its emission within the range of 1500-2100 nm, in both water and ethanol (Figures 1b and 1c, respectively). The emission spectra of VGE in ethanol revealed two emission peaks, at 1743 nm and 1844 nm.

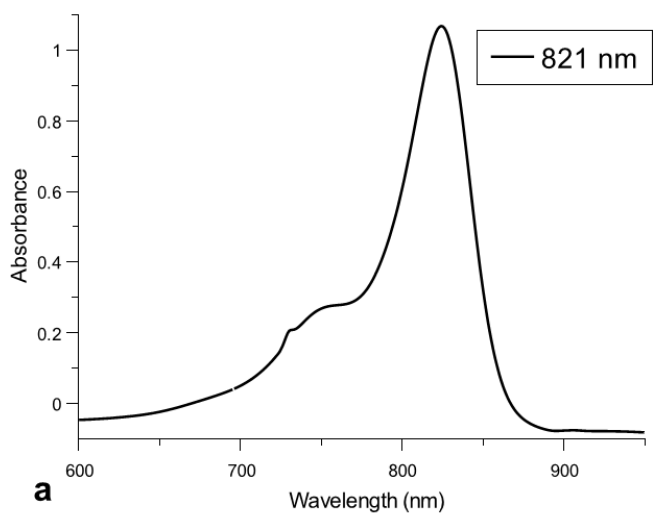

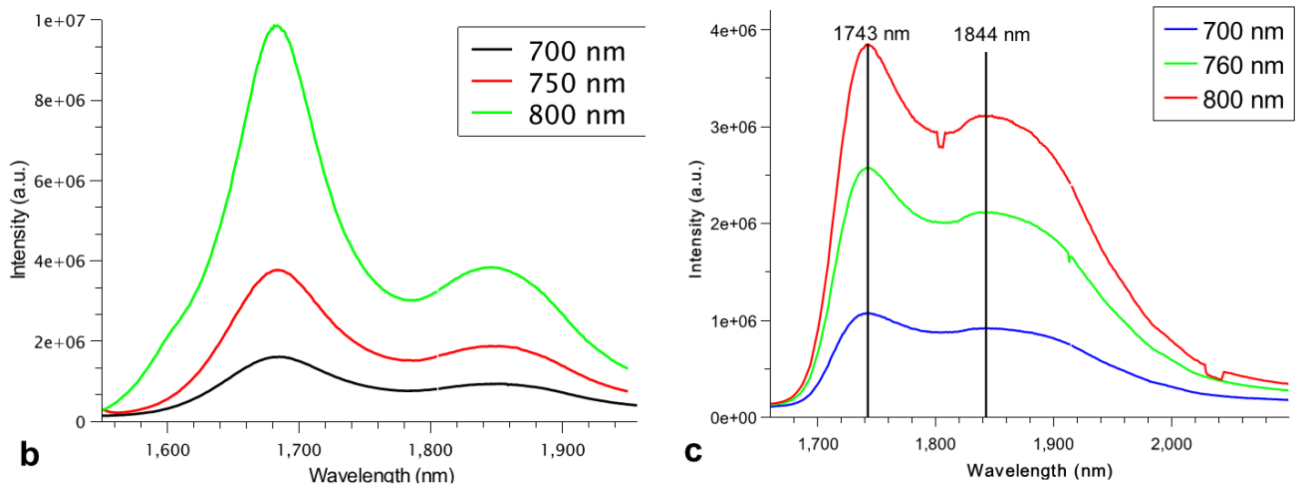

**Figure 1**: Spectral properties of VGE in Ethanol and Water. (a) Absorption spectra of VGE 25 μM in ethanol. Emission spectra of VGE 23 μM in water (b) and in ethanol (c). Legends show the excitation wavelengths in each experiment.

## Absorption of CGC and NCD

The excitation wavelength range for CGC and NCD dyes was studied to determine their emission properties. We examined the fluorescence spectra of these tricarbocyanines by exciting the dyes across their entire absorption spectra. Figure 2 shows the absorption spectra of CGC and NCD in ethanol and water, resembling to those obtained in methanol, except for the shoulder peak at 730 nm[25, 26]. The absorption spectrum of NCD exhibited a peak at 802 nm in water and a peak at 811 nm in ethanol, with shoulders at 730 nm for both water and ethanol solution (Figure 2a). For CGC, the absorption peaks are at 786 nm for ethanol and 775 nm for water solution, both with a shoulder at 730 nm for ethanol and 706 nm for water (Figure 2b). This discrepancy elucidates why the green CGC ethanolic solution transitions to an azure upon dilution in water.

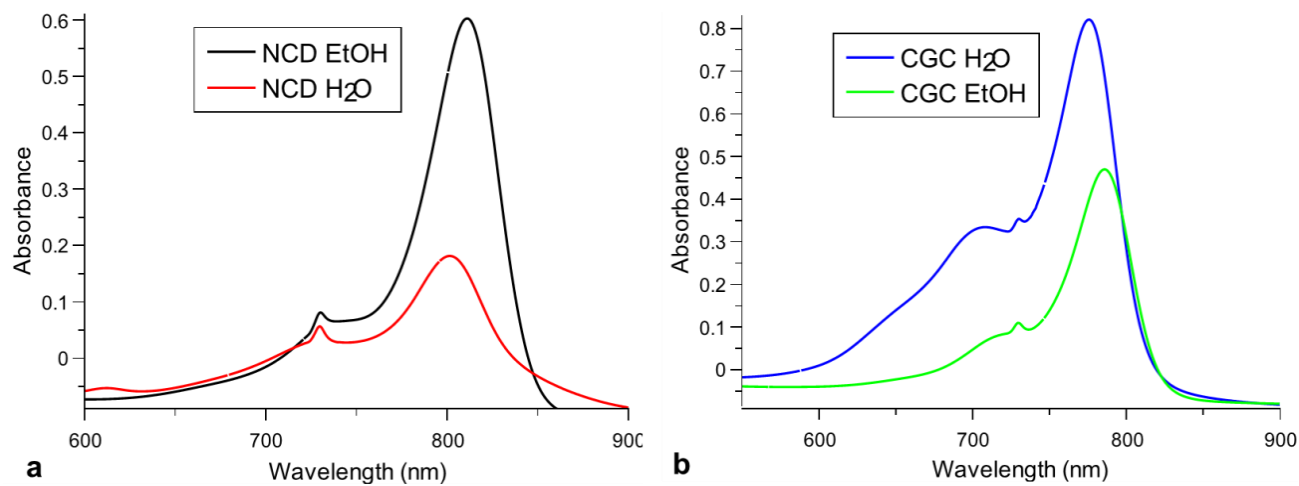

**Figure 2:** Absorption spectra of (a) 15 μM NCD in water and ethanol and (b) CGC 130 μM in water and 13 μM ethanol.

**Emission spectra of CGC and NCD**

Figure 3 shows the emission spectra of NCD and CGC in ethanol and water. We chose the excitation wavelengths to be between 630-850 nm for NCD and 600-840 nm for CGC, to cover most of their absorption range (presented in Figure 2 above). Emission spectra were recorded at 1500-2100 nm, in both ethanol and water. Both dyes show a double peak structure, when the longer emission peaks are in the 1800 nm zone and the shorter is at the 1600 nm zone. A large s shift, of ~1000 nm, similar to other organic molecules[16, 27], was also observed, with a tail emission up to ~2200 nm (Figure 4). The dyes exhibit emission at a constant longer wavelength, across a broad spectrum of excitation wavelengths, showing independence from the excitation wavelength.

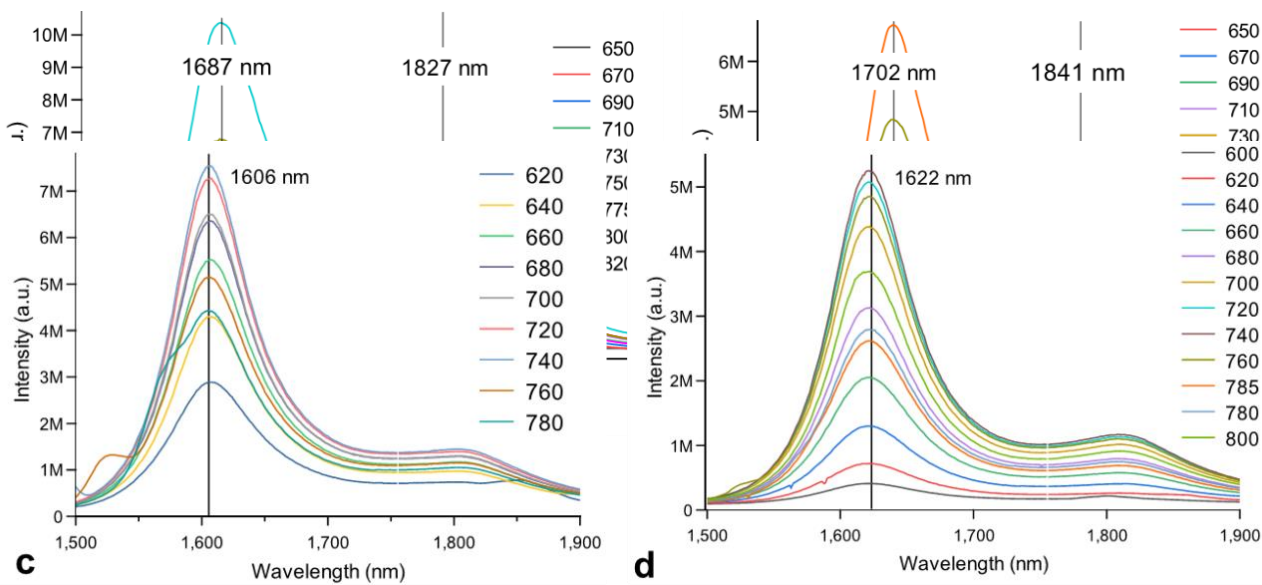

**Figure 3:** Emission of NCD 30 μM in water (a) and 15 μM in ethanol (b). Emission of CGC 25 μM in water (c) 13 μM ethanol (d). Legends show the excitation wavelengths in each experiment, corresponding with the different colors of the curves.

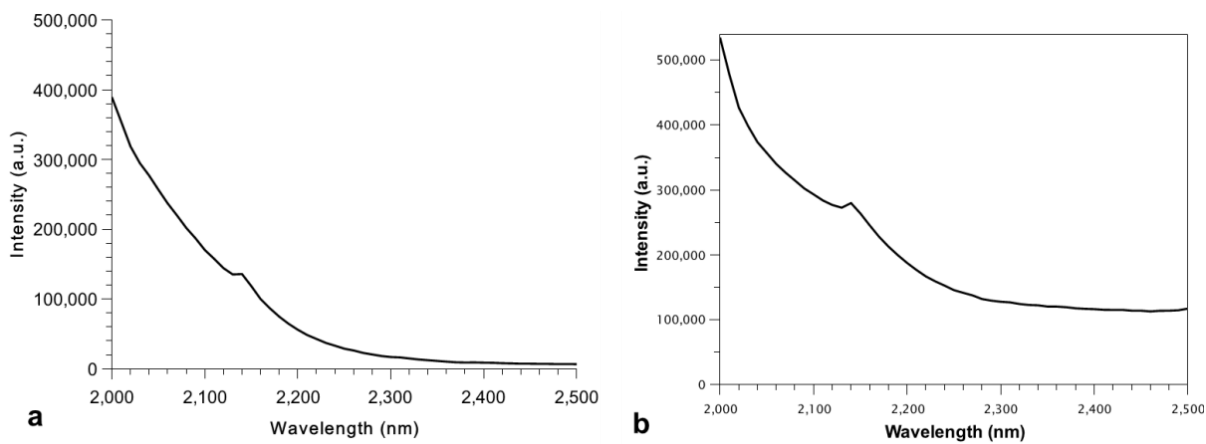

Figure 4: Tail fluorescence emission of 635 μM CGC (a) and 15 μM NCD (b) in ethanol.

To confirm that the emission bands originate from the cyanine dyes themselves, we measured spectra of ethanol in the LWNIR range, where only 1.5 and 2nd order emissions are observed (Supporting Information, figure S0 a and b). We also recorded the emission spectra covering the known emission range. One can observe both emissions when the known emission bands at 820 and 910 nm in CGC and 852 and 920 nm in NCD were observed. The intensities of the new peaks are 2-6 times lower than those of the known emission bands, but still distinct, especially the 1600~ band (Supporting Information, Figure S0 c and d). Since the emission bands at 800~ and 900~ are known and lie in the short-wavelength NIR, we focus on the new LWNIR emission bands.

The new emission bands of NDC and CGC at 1500-2100 nm should arise from a different energy level, with highly allowed transition, in these tricarbocyanines have not yet been explained. Further theoretical studies on the exact orbital arrangement of this excited state should be carried out in the future. It is also observed that the reproducibility of the structure and emission spectra is high and independent of the excitation wavelength. **Error! Bookmark not defined.**Additionally, the emission maxima of our tricarbocyanines in ethanol were ~20 nm longer than the corresponding emissions in water, indicates a solvatochromic effect that modifies the energy levels of the dyes. A previous study has also shown that two tricarbocyanines, excited in alcoholic solvents suggested emission peaks' ratio dependance on the polarity of the solvent[28].

**Photostability of CGC and NCD in aqueous solutions.**

To determine the optimal conditions for observing the emission spectra of the dyes, we investigated the stability of CGC and NCD emissions in both ethanolic and aqueous solutions under different experimental conditions.. Photostability analysis under continuous light irradiation with 100 watt tungsten lamp, reveals degradation of both 25 μM CGC and 30 μM NCD in water, accompanied by a small (3-5 nm) shift in the prominent peak

(Figures 5a and 5b). Notably, while both peaks of CGC degrade proportionately, NCD's prominent peak exhibits slower degradation compared to its second emission peak at 1825 nm (figure 5a). Degradation of NCD and CGC in water, as observed in Figures 6a and 6b, is common for cyanine dyes and was and well known.[26,25, 29] In ethanol, however, both CGC and NCD exhibit stable emission spectrum intensities (Figures 5c and 5d).

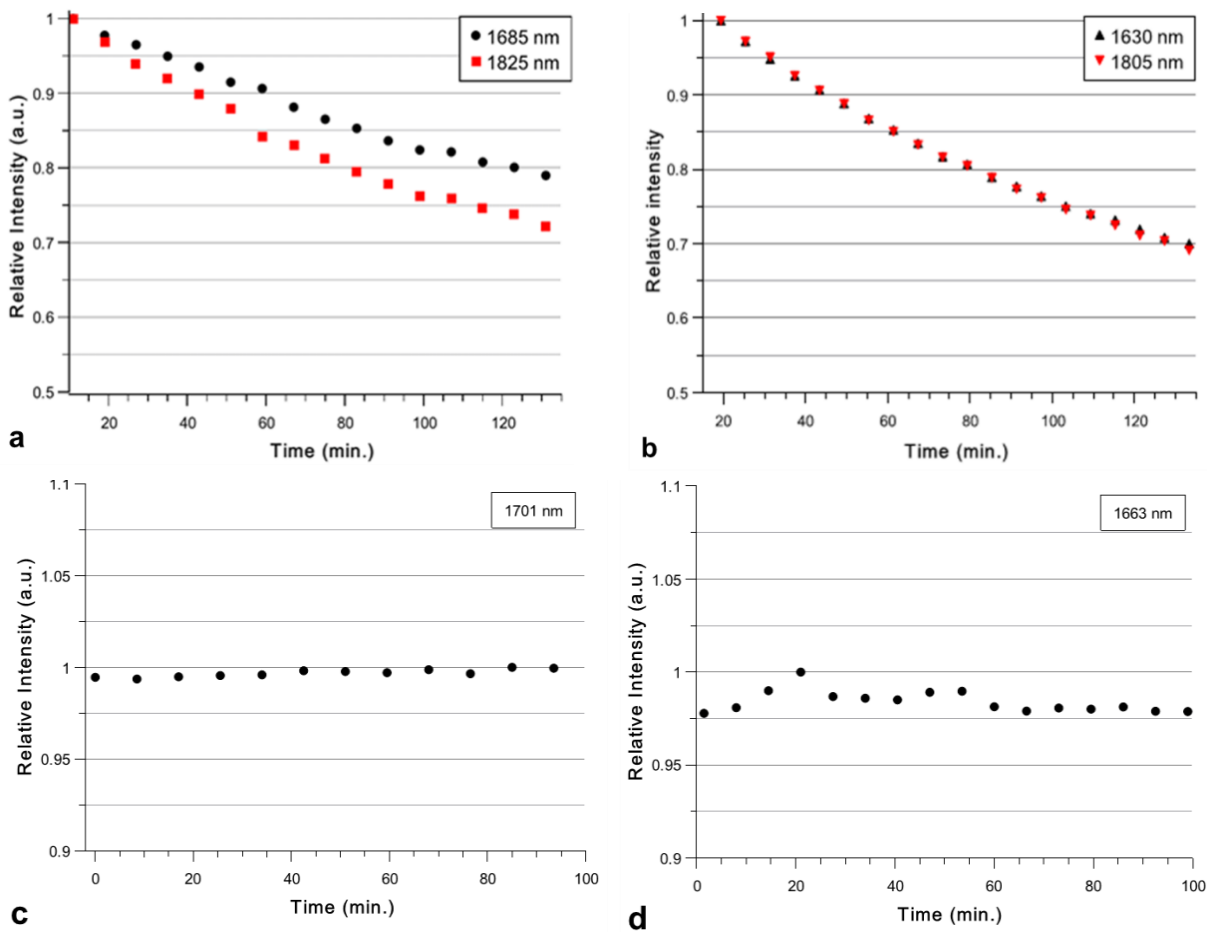

**Figure 5:** Photostability tests for NDC and CGC. Intensity of the peak/s versus time for 30 μM NCD (a) and 25 μM CGC (b) in water and 10 μM NCD (c) and 25μM CGC (d) in ethanol. The legends in all panels indicate the emission peaks that were studied.

## Stability of CGC and NCD in Water in Dark, and upon Heating

Noting the decomposition of CGC and NCD in water under irradiation, we endeavored to determine their stability in water under dark conditions. (*Supporting Information*, Figure S1) (a and b) illustrates the stability of 15 μM NCD and 25 μM CGC in water in darkness, where their emission peaks remained consistent even after more than 2 hours. We also measured their fluorescence upon heating it in opaque vials submerged in water, followed by sampling and measurement at ambient temperature. (*Supporting Information*, figure S1 (c and d)) demonstrates the stability of 7 μM NCD and 13 μM CGC after heating to $45^0$C, suggesting their suitability for use in biological experiments. Additional measurements revealed that the fluorescence intensity of both NCD and CGC in ethanol increases during the first hour until stabilizing, while the peak patterns remained consistent across experiments (*Supporting Information*, Figure S2). We also measured the emission spectra of 15 μM NCD (a) and 25 μM CGC (b) in various pH buffers, as well as in water solution (control solution). Despite variations in intensity, emission spectra persist across pH 4, 7, and 10 buffer solutions, with a slight shift of the main peak, which is due to the different environment of the buffer solution around the dye.. Our findings suggest stability of CGC and NCD within the pH range of 4 to 10 (*Supporting Information*, Figure S3).

**Concentration Effect on the Position of the Peaks**

A careful examination of the experiments described above suggests that the emission peak positions of CGC and NCD are not entirely constant. This phenomenon was more pronounced in CGC, where its main peak varied from 1606 nm to 1633 nm in water and from 1622 nm to 1663 nm in ethanol (see Figures 3c and 3d and *Supporting Information,* Figure S4). It is well known that the higher is the dye's concentration, the more likely is to form aggregates, which are characterized by a spectral red shift. However, aggregates have a specific emission pattern[1, 2], which do not match the LWNIR peaks presented in this paper. In addition, aggregates are not prone to solvatochromism, while our results clearly show the presence of this effect. Consequently, we further explored the effects of solvatochromism on both CGC and NCD by using a range of concentrations in ethanol and water, when the ethanolic solutions were diluted from the CGC stock solution, while the aqueous solutions were prepared by evaporating the ethanolic CGC solution and then dissolving with distilled water. Figures 6 a and b show two normalized emission CGC peaks of the dye versus its concentration. The lower energy peak is only pronounced at concentrations higher than 0.1 μM. As the concentration decreases, a higher energy peak, associated with the monomeric form, becomes evident. The ratio between the low and the high energy peaks is demonstrated in Figures 6c and 6d. Although its intensity decreases, the low-energy peak remains detectable even at low concentrations.

Figures 6e and 6f show the shift of CGC two peaks versus its concentration in ethanol and water, respectively. Initially, decreasing the concentration of the dye caused a blue shift. When the concentration was further reduced to the micromolar range, it reached constant emission bands, of 1612 nm and 1590 nm, while the other peak reaches a constant emission at 1805 and 1800~ nm for water and ethanol solutions, respectively. Also, CGC emission in water shows much lower energy with emission peak at 1872 nm, which is one of the lowest known emissions for organic molecules. Complementary experiments have shown that NCD emissions versus the concentration of the dye show similar trend (*Supporting Information*, Figure S5)

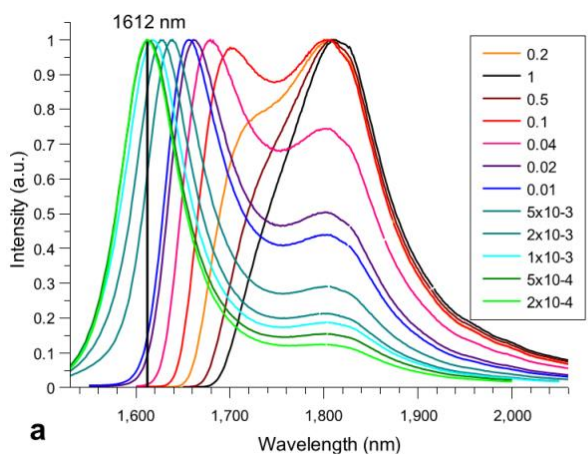

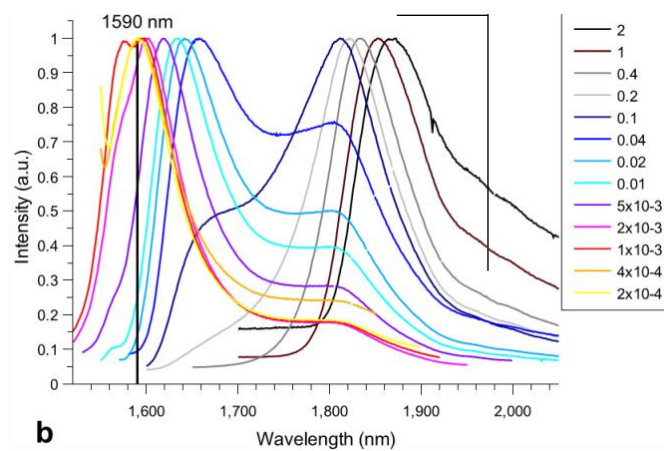

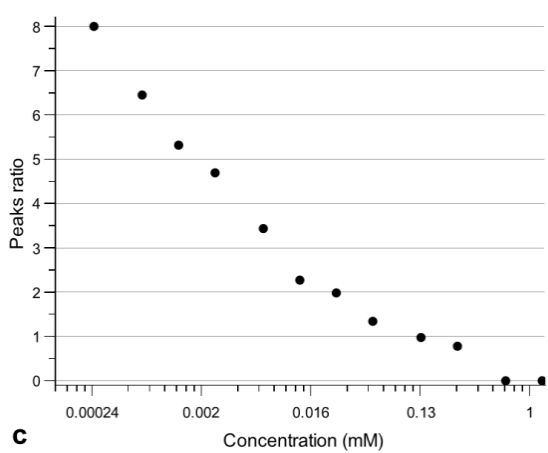

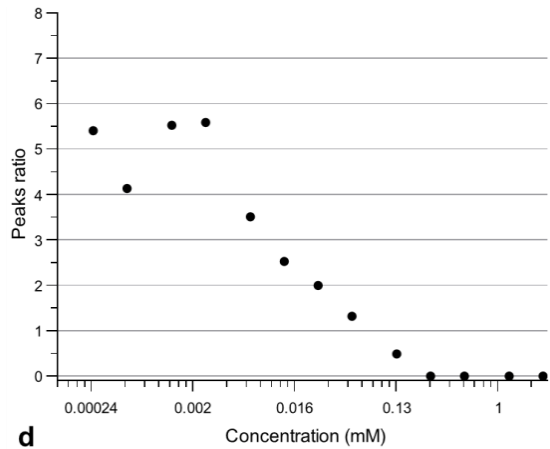

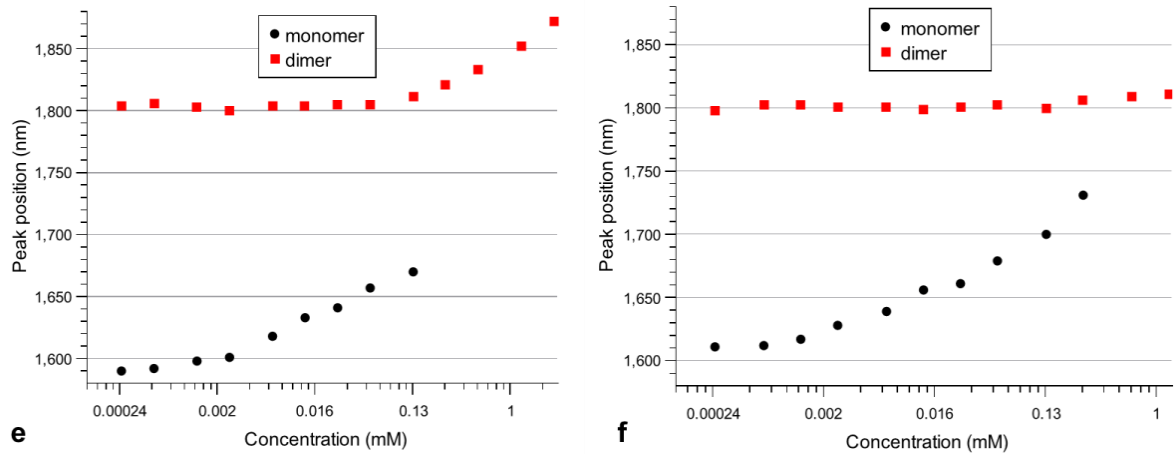

**Figure 6**: Emission of CGC in a different concentration range, (a) in ethanol and (b) in water. Intensities were normalized for clarification. Legends show the relative concentration of the CGC when 1 is the concentration of the stock solution (1.27 mM). Dependence of the peak's ratio on the concentration of CGC in ethanol (c) and in water (d). Scale is logarithmic. Dependence of the peak's position on the concentration of CGC in ethanol (e) and in water (f) scale is logarithmic. The numbers in the legend indicates the dilution factor from the stock solution of the CGC (1.27 μM).

## Dynamic Light Scattering (DLS), NMR Quantum yield and Fluorescence lifetime (FLT) measurements

DLS measurements were conducted for VGE in ethanol, and for CGC and NCD in both ethanol and water. No observable aggregates were detected, even when high concentrations of stock solution were measured (*Supporting Information*, Figure S6). FLT of CGC and NCD in ethanol were 0.19 and 0.55 ns respectively, while shorter FLT were observed for CGC and NCD in water, of 0.06 and 0.35 ns. unfortunately quantum yields calculations were unsuccessful.

We also tried to observe the structural differences between the low and high concentration of CGC solutions. The $^1$H NMR spectra of 13 μM and 1.27 mM CGC in $D_2O$ are shown in (Supporting Information, Figure S7). Although the NMR spectrum of the high concentration is similar to a previous report[30], unfortunately, no cyanine spectrum was observed for the low concentration sample, which was below the detection limit of the instrument. Therefore, we cannot compare the structural differences between the low and high concentration solutions.

The present work shows the fluorescence emission of the cyanine dyes VGE, NCD and CGC in the LWNIR range. The molecular structure of these molecules suggests that their long cyanine chain causes the LWNIR emission. We show, for the first time, a unique behavior of the IR 820, IR 806, and IR 783, with emission in the LWNIR range up to 2200 nm. The emission bands do not depend on the excitation wavelength but can vary between 1605-1845 nm, according to the solvent and pH, with a fluorescence emission tail up to ~2200 nm. The emission peaks of CGC and NCD, which are similar to those of VGE, are larger than 1500 nm, presenting Stokes shifts greater than 1000 nm ($>7000$ cm$^{-1}$). We also tested the stability of CGC and NCD under different conditions and found them to be stable at pH values of 4-10 and up to 45°C. The dyes showed photostability in ethanol, whereas in water they were only stable in the dark. We have also shown the peak positions and the relationship between the peak intensity is concentration dependent when the higher the concentration of the cyanine the more prominent is the low energy peak and vice versa. The DLS measurements indicated no significant aggregation, and the presence of solvatochromism suggests that the dyes remain in their monomeric form (Figures 3 and 6. Additionally, results suggest a significant concentration-dependent effects on the peaks Figure 6 but a minimal impact of the pH on the peak positions (Figure 5). Based on this data, we propose that conformational changes may cause the energy level shift between the two peaks, and that this shift is concentration-dependent in the tricarbocyanine dye. The peaks may also arise from intramolecular charge transfer processes[31]. Further studies at different experimental conditions on different dyes may enlighten more information about the origin of these new bands in the LWNIR.

**Conclusions**

The novel LWNIR fluorescence bands in tricarbocyanine dyes open up exciting possibilities for advanced biomedical applications. These dyes could significantly improve in vivo imaging, allowing deeper tissue penetration due to the reduced scattering and absorption of LWNIR light in biological tissues. This makes them ideal candidates for high-resolution imaging in areas such as cancer diagnostics, brain imaging, and molecular tracking. Additionally, their fluorescence properties can enhance phototherapy techniques, including photothermal and photodynamic therapies, by providing better targeting and visualization of treatment areas. Their stability under varying conditions, such as pH and temperature, further expands their potential use in drug delivery systems and biosensing, enabling more effective and real-time monitoring of biological processes.

## Supporting Information

The following information is available free of charge.

Emission spectra of CGC in NCD in buffer solutions; emission intensities of CGC and NCD in water under dark conditions, emission intensities of CGC and NCD after heating, emission intensities of NCD and CGC in time and emission of NCD at different concentrations; Intensities of CGC in ethanol at different slit width; DLS spectra of CGC and NCD solutions; NMR spectra of CGC solutions.

## Author Information

### Corresponding Author
Rinat Ankri: rinatsel@ariel.ac.il

### Authors
Nitzan Dar - Department of Physics, Faculty of Natural Science, Ariel University, Ariel 40700, Israel

Rinat Ankri - Department of Physics, Faculty of Natural Science, Ariel University, Ariel 40700, Israel; https://orcid.org/0000-0002-7686-9605;

## Acknowledgements
The work was supported by the Council for Higher Education, Israel.